\documentclass{pnastwo}

\usepackage[pdftex]{graphicx}

\usepackage{amssymb,amsfonts,amsmath}
\usepackage{color}

\usepackage{cite}

\pdfpageattr{/Group <</S /Transparency /I true /CS /DeviceRGB>>}

\newcommand\bF{{\mathbf{F}}}
\newcommand\bp{{\mathbf{p}}}
\newcommand\bP{{\mathbf{P}}}

\newcommand\bx{{\mathbf{x}}}
\newcommand\bu{{\mathbf{u}}}
\newcommand\dhat{{{\mathbf{d}}}}
\newcommand\zhat{{\mathbf{e}_z}}
\newcommand\thetahat{{\mathbf{e}_\theta}}
\newcommand\tI{{\mathbb{I}}}

\contributor{Published in Proceedings
of the National Academy of Sciences of the United States of America}
\url{www.pnas.org/cgi/doi/10.1073/pnas.1302736110}
\copyrightyear{2013}
\issuedate{August 27, 2013}
\volume{110}
\issuenumber{35}

\begin{document}



\title{Cytoplasmic streaming in plant cells emerges naturally by microfilament self-organization}





\author{Francis G. Woodhouse\affil{1}{Department of Applied Mathematics and Theoretical Physics, University of Cambridge, Wilberforce Road, Cambridge CB3 0WA, UK}
\and
Raymond E. Goldstein\affil{1}{}}

\contributor{Published in Proceedings of the National Academy of Sciences
of the United States of America}

\maketitle

\begin{article}

\begin{abstract}
Many cells exhibit large-scale active circulation of their entire fluid contents, a process termed cytoplasmic streaming. This phenomenon is particularly prevalent in plant cells, often presenting strikingly regimented flow patterns. The driving mechanism in such cells is known: myosin-coated organelles entrain cytoplasm as they process along actin filament bundles fixed at the periphery. Still unknown, however, is the developmental process which constructs the well-ordered actin configurations required for coherent cell-scale flow. Previous experimental works on streaming regeneration in cells of Characean algae, whose longitudinal flow is perhaps the most regimented of all, hint at an autonomous process of microfilament self-organization driving the formation of streaming patterns during morphogenesis. Working from first principles, we propose a robust model of streaming emergence that combines motor dynamics with both micro- and macroscopic hydrodynamics to explain how several independent processes, each ineffectual on its own, can reinforce to ultimately develop the patterns of streaming observed in the Characeae and other streaming species.
\end{abstract}



\abbreviations{IZ, indifferent zone}



\dropcap{C}ytoplasmic streaming pervades the vast spectrum of cell types, achieving great diversity in order to fulfill many 
varied goals. {\it Drosophila} oocytes use disorderly streaming to localize proteins without counter-productive 
backflows~\cite{KhucTrong2012,Ganguly2012}; {\it C. elegans} couples cortical flow with bistable pattern formation for 
polarity determination~\cite{Goehring2011}; {\it Elodea} leaf cells circulate chloroplasts in response to illumination~\cite{Allen1978}; 
and pollen tubes transport tip growth material in fountain-like flows~\cite{Taylor1997,Hepler2001}, to name four disparate examples.  
At their core, either as the end goal or within the mechanism itself, lies the biological pillar of pattern 
formation~\cite{Howard2011}.

The different realizations of streaming can be effectively categorized by the degree of order inherent in the 
observed flow~\cite{Kamiya1959,Allen1978}. The most disordered instances often function as part of 
cellular patterning processes, as in {\it Drosophila}. In contrast, the orderly types must themselves already 
be the product of some pattern-forming process in order to exist at all. The basic paradigm that underlies 
streaming, motor proteins interacting with polymer filaments, has been seen to possess many pattern forming 
behaviors in both theoretical~\cite{Kruse2004, Allard2010, Surrey2001} and 
experimental~\cite{Surrey2001, Narayan2007, Schaller2010, Sanchez2012} settings,
with recent work beginning to incorporate the effects of cylindrical cell-like domains~\cite{Zumdieck2005, Srivastava2013}. However, such studies are 
often `bottom up', taken out of context from concrete biological systems, and in particular no direct connection 
has been made to the development of cytoplasmic streaming.

\begin{figure}
\centering
\includegraphics[width=\columnwidth]{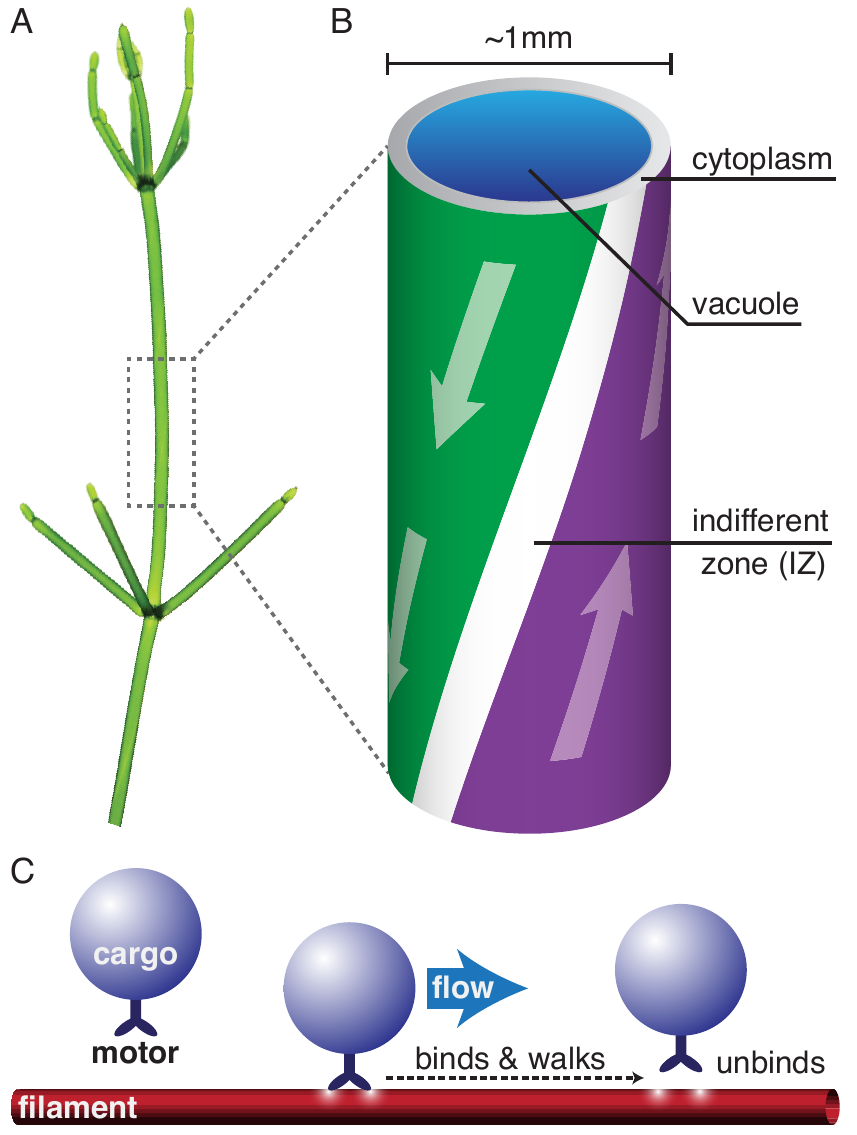}
\caption{Cytoplasmic streaming in {\it Chara corallina}. (A) Internodes and branchlets of {\it Chara}. Individual internodes can grow up to 10~cm long. (B) Rotational streaming in a single internodal cell of {\it Chara}. The stripes indicate the polarity of actin cables at the periphery driving flow in the cytoplasm and vacuole. (C) Microscopic mechanism driving plant cell streaming. Cargo-carrying myosin motors bind to actin filaments and entrain flow as they walk.\label{fig:1}}
\end{figure}

If we wish to understand what fundamental dynamics drive the formation of orderly streaming and connect the microscopic 
with the macroscopic, an alternative `top down' approach is warranted. To this end, we approach the problem by way 
of a concrete, prototype system. We adopt perhaps the most strikingly patterned example of all, the aquatic 
alga {\it Chara corallina} (Fig.~\ref{fig:1}A). The giant cylindrical internodal cells of {\it Chara} measure 1~mm in diameter and up to 10~cm 
in length. First observed in 1774~\cite{Corti1774}, its rotational streaming---termed \emph{cyclosis}---is driven by 
vesicles (in the endoplasmic reticulum) coated with the motor protein myosin~\cite{Nothnagel1982, Kachar1988, Williamson1993} 
sliding along two oppositely-directed longitudinal stripes of many continuous, parallel actin filament 
cables~\cite{Sheetz1983, Shimmen2007} (Fig.~\ref{fig:1}B,C). Every cable is a bundle of many individual actin filaments, each possessing the same intrinsic polarity~\cite{Kersey1976a}; myosin motors walk on a filament in a directed fashion, from its minus (pointed) end to its plus (barbed) end. These cables are bound to the 
cortically-fixed chloroplasts at the cell periphery~\cite{Kersey1976b} in a `barber pole' twist, generating flow speeds of 
50--100~$\mathrm{\mu m / s}$~\cite{Goldstein2008,vandeMeent2008,vandeMeent2010}. It is unclear how this simple yet striking pattern 
forms during morphogenesis; what process is responsible for forming two, and only two, precisely equal stripes? One could 
imagine it to be the result of complex chemical pre-patterning, but experimental evidence suggests otherwise.

\begin{figure}[t]
\centering
\includegraphics[width=\columnwidth]{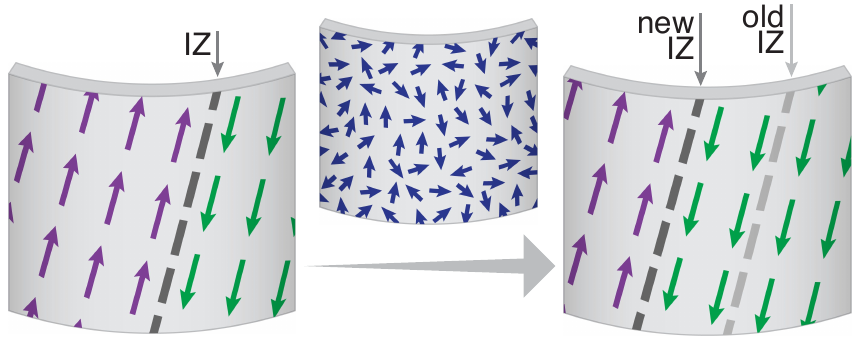}
\caption{Disruption, rearrangement and resumption of streaming in {\it Chara}~\cite{Foissner2000}. Filament bundles disorganize and reform with the indifferent zone in a new location.\label{fig:2}}
\end{figure}

Foissner {\it et al.}~\cite{Foissner2000} introduced the agents cytochalasin-D and oryzalin into young {\it Chara} cells 
in order to arrest streaming and completely disorganize the established actin cables. They found a gradual 
recovery process (Fig.~\ref{fig:2}) whereby streaming initially resumes in a disordered saltatory fashion. Over time
the filaments reorganize, first into locally ordered `streamlets', until eventually the initial filament layout is re-established.
They note one crucial difference from the pre-arrest configuration: the so-called indifferent zone (IZ; Fig.~\ref{fig:1}B) 
separating the up- and down-streaming stripes, initially identifiable by a line of missing chloroplasts, is often found in a new position. Put another way,  
the entire pattern has shifted around the central axis. Together with the local-to-global progression, also observed in other 
cell types~\cite{Jarosch1956,Kamiya1959}, a self-organization process is implicated.

Similar behavior has been 
induced in rotationally streaming {\it Vallisneria} leaf cells where flow handedness was preserved only if some `seed' 
filament bundles survived disintegration~\cite{Ryu1995}, a hallmark of a symmetry-breaking self-organization process. 
Healing of localized wounds in Characean cells also furnishes evidence, where the actin network regenerates in a 
gradually organizing process~\cite{Foissner1996, Foissner1997}. Even {\it ex vivo} experiments lend credence to such a
theory: cytoplasmic droplets forcibly extracted from {\it Chara} cells can begin spontaneously circulating 
at a steady rate~\cite{Yotsuyanagi1953a, Yotsuyanagi1953b, Kamiya1959} likely owing to a self-organizing 
process~\cite{Woodhouse2012b}. Furthermore, images of filament networks in very young Characean cells
show long cables forming only after elongation begins, with just a 
disorderly meshwork at genesis~\cite{Wasteneys1996}.

In this work, we will unify this wealth of experimental evidence for a self-organization mechanism by constructing a simple theoretical model of streaming emergence during cell development.
The model will couple the hydrodynamics of the cytoplasm with the dynamics of a microfilament suspension
subject to a few key reorienting and organizing effects. We will then illustrate the model's capacity for mimicking the emergence of Characean streaming by way of an example and a parameter space scan to identify
regimes of robust cyclosis development. Finally we will discuss the model in the context of pathologies
and disruptions to understand the importance of each of the components of the filament dynamics.

\section{Model}

A young {\it Chara} cell is modeled as a cylinder of radius $R$ with periodic boundaries at a distance $L$ apart. The fixed 
chloroplasts lie at the edge, with the subcortical cytoplasm a thin cylindrical layer between the chloroplasts and the 
vacuolar membrane; the vacuole then comprises the bulk of the cell. The cytoplasm is taken to consist of a layer of 
short, initially disordered actin filaments beneath a layer of myosin-coated endoplasmic reticulum vesicles, which drive flow by forcing the cytoplasm as they process on the actin filaments (Fig.~\ref{fig:1}C). Because the subcortical layer is very thin compared 
to the vacuole, we approximate the cytoplasmic layer as a purely two-dimensional cylindrical shell. We also assume
that the effective viscosity contrast between cytoplasm and vacuolar fluid is sufficiently large to take vacuolar flow as 
purely passive, induced by cytoplasmic flow~\cite{Goldstein2008} but not affecting it. This yields a truly two-dimensional
problem. 

Our final simplification is to neglect the helical twist in the streaming pattern (Fig.~\ref{fig:1}B).
No discernible twist exists in young cells
with developed actin cables~\cite{Foissner2000}, so its appearance is more likely connected to later development of the microtubule
cytoskeleton~\cite{Green1962} by a process of twisting growth once the main cables are established and bound
to the cortical chloroplasts~\cite{Green1954}. The twist appears to be advantageous primarily for large, mature cells,
where it enhances vacuolar mixing and nutrient uptake~\cite{Goldstein2008,vandeMeent2008}.

We will now formalize the hydrodynamics. Endow the shell with cylindrical coordinates $(\theta,z)$ and corresponding unit vectors $\thetahat,\zhat$. Let 
$\bu(\bx)$ be the velocity field of the suspending cytoplasmic fluid. The small velocity of streaming 
allows us to work in the zero Reynolds number limit. We then take $\bu$ to obey the forced Stokes equations with friction,
\begin{align*}
-\mu \nabla^2\bu + \nu \bu + \nabla \Pi = \bF~,
\end{align*}
where $\Pi(\bx)$ is the two-dimensional pressure and $\bF(\bx)$ is the filament-induced forcing, discussed shortly.  
The frictional term $\nu \bu$ captures the effect of the no-slip boundary proximity in our thin-shell approximation, in a 
manner similar to Hele-Shaw flow (SI Text).

There are two crucial constraints on the flow field that must be incorporated. First, because the tonoplast 
membrane, which separates the cytoplasm and vacuole, behaves as a two-dimensional incompressible 
fluid~\cite{Woodhouse2012a}, we must have \emph{incompressibility} $\nabla \cdot \bu = 0$. Second, 
the presence of end caps on the cell must be acknowledged, either through explicit modeling of the end geometry or, 
as we do here, qualitatively through the simple extra constraint of \emph{zero net flux} 
$\int_0^{2\pi} \int_0^L \bu\cdot\zhat \, dz d\theta = 0$, essentially due to the incompressibility of a finite domain in the $\zhat$ direction. This is balanced by allowing a longitudinal pressure gradient 
$\Pi_0$ and writing $\Pi(\bx) = \Pi_0 z  + \Pi'(\bx)$, with $\Pi'(\bx)$ the remaining fully periodic pressure field.

We now turn to the filament suspension, whose dynamics will incorporate several crucial effects.
The filaments are taken to be \emph{restricted}: passive advection and 
shear alignment in the flow $\bu$ is inhibited by a factor $\epsilon$ due to frictional or binding effects of the filaments 
on the chloroplasts or with cortical polymer networks.
As a consequence of restriction, the filaments are
\emph{non-self-advective}: a vesicle walking forwards on a filament will 
induce only a negligible backward propulsion of the filament itself.
The filaments are also taken to \emph{spontaneously bundle}: filaments will 
locally align with each other, controlled by a coupling constant $\alpha_p$ (a rate constant for the exponential growth of small local polarization), mimicking the presence of bundling proteins.
(Inhibiting the action of bundling proteins thus corresponds to setting $\alpha_p = 0$.)

\begin{figure}
\centering
\includegraphics[width=\columnwidth]{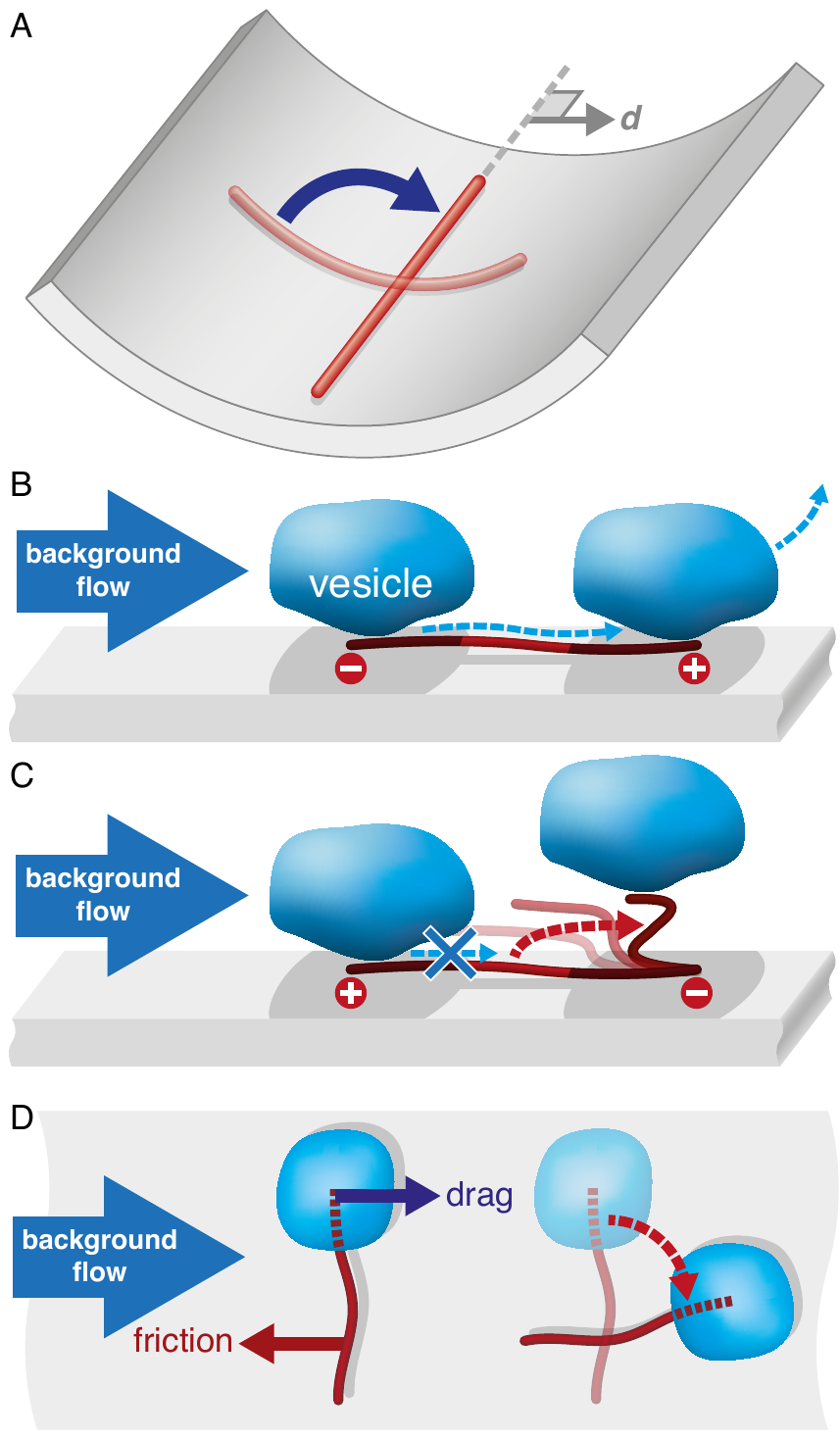}
\caption{Filament reorganization mechanisms. (A) Repulsive direction: a filament preferentially reorients away from a direction
$\dhat$ either to minimize physical curvature, or minimize interactions. (B-D) Polar flow reorientation: rather than aligning nematically with flow shear, filaments can align in a polar fashion with flow itself. Alignment with the plus end downstream is stable (B), but hydrodynamic drag on vesicles bound to the filament competes with friction from wall restriction when the motor is not free to walk downstream (C,D).
\label{fig:3}}
\end{figure}

To represent non-spherical cellular geometry or substrate patterning we include a \emph{repulsive direction} 
$\dhat$ which filaments will preferentially avoid, with coupling constant $\kappa$ (Fig.~\ref{fig:3}A); this will be set here to 
$\dhat = \thetahat$, the circumferential direction. While apparently a strong assumption, there is remarkable experimental precedent for such an 
effect whereby filaments can reorient circumferentially around cells upon inhibition of
binding or director components \cite{Masuda1991,Foissner2002,Tominaga1997}.

In the absence of any other contributing factors, the effects described above will tend to produce a uniformly polarized  
suspension in a direction perpendicular to $\dhat$. There is one more key factor: \emph{polar flow alignment}, with 
coupling constant $\alpha_u$ (encoding the linear growth rate of filament polarization with cytoplasm flow, similar to $\alpha_p$). We 
argue that the presence of myosin-coated vesicles brings about alignment of the filaments with the flow field $\bu$ itself in a truly polar fashion, minus end upstream and plus end downstream, supplementing traditional nematic alignment with flow shear $\nabla\bu$. 
Suppose that a motor-laden vesicle approaches a filament, advected by a net background flow $\bu$ from the action of other vesicles processing along filaments nearby.
There are then three basic cases to consider for the filament orientation relative to the flow.
If the vesicle approaches a microfilament oriented parallel with $\bu$---that is, minus end upstream so the myosin walk direction is with the flow---then the vesicle can bind, process along the filament at a speed $\sim |\bu|$,
and unbind with little change to the filament (Fig.~\ref{fig:3}B). This implies parallel flow alignment is stable.
Anti-parallel alignment, with plus end upstream and walk direction against the flow, is unstable: a vesicle bound upstream on the filament will be impeded from walking by hydrodynamic drag, stall on the restricted filament, and twist it into the flow (Fig.~\ref{fig:3}C).
Finally, perpendicular alignment induces reorientation either parallel or anti-parallel to the flow upon binding of a vesicle, because a torque is generated 
by the competition of filament restriction against hydrodynamic drag (Fig.~\ref{fig:3}D).
This effect is not unique to microfilaments: thin-film dumbbell microswimmers are subject to the same
polar alignment, driven by asymmetric friction forces \cite{Brotto2013}.

For this reorientation mechanism to occur and be robust (that is, stable when oriented parallel with the flow), there are two important conditions. First, the timescale $\tau_r$ for filament reorientation must be shorter than the timescale $\tau_d$ for motor detachment after stalling \cite{Parmeggiani2001} in order for any appreciable filament reorientation to occur in the flow-antiparallel and flow-perpendicular cases. Second, the timescale $\tau_p$ for a motor to process along a flow-parallel filament must be shorter than the filament reorientation timescale $\tau_r$, in order for orientation parallel with the flow to be stable; if this is not the case, then no orientation is stable, and the filament will writhe continuously over time irrespective of its orientation relative to the background flow. Thus, we require $\tau_p \ll \tau_r \ll \tau_d$. The sliding speed $\sim 100\,\mathrm{\mu m / s}$ \cite{Shimmen2004} on $\sim 10\,\mathrm{\mu m}$ filaments yields a quick procession timescale $\tau_p \sim 10^{-1}\,\mathrm{s}$. By contrast, it has been suggested that the very fast streaming may in fact be due to the action of many non-processive motors on one vesicle acting simultaneously \cite{Kimura2003,Shimmen2004} which would likely yield a long stall detachment time $\tau_d > 1\,\mathrm{s}$ due to the long dwell time of individual characean myosins \cite{Kimura2003}. Microfluidic studies indicate that the reorientation timescale $\tau_r$ lies in-between these two extremes \cite{Kantsler2012}, indicating that this reorientation mechanism is indeed viable.

To describe the microfilaments we use a formalism from the theory of `active suspensions'
\cite{Saintillan2008, Woodhouse2012b}. The full derivation of the dynamics can be found in the SI Text; we summarize it here, as follows. At each point $\bx$ on the cylinder, 
define the angular distribution function $\Psi(\bx,\bp)$ of microfilaments with unit direction $\bp$. From this, we 
define the local filament concentration $c(\bx) = \int_\bp \Psi(\bx) d\bp$ and average orientation director 
$\bP(\bx) = c(\bx)^{-1} \int_\bp \bp \Psi(\bx) d\bp$. The distribution $\Psi(\bx,\bp)$ obeys a Smoluchowski equation, 
whose orientational moments yield the dynamics of the concentration and director fields after 
applying a simple moment closure approximation, whose viability is discussed further in the SI Text. Our non-self-advective supposition then reduces the
concentration dynamics to pure advection-diffusion (SI Text), allowing us to neglect fluctuations and 
reduce to a \emph{uniform concentration} $c(\bx) \equiv c_0$.

The system is closed by defining the flow forcing $\bF(\bx)$. As the filaments are restricted by frictional forces with the outer wall, the primary contribution 
of a vesicle processing along a filament is a point force (i.e. a Stokeslet), suggesting a simple relationship 
$\bF \propto \bP$. However, global streaming does not persist in the presence of filament unbundling agents, while 
the actin-myosin interaction remains uninhibited~\cite{Foissner2000}. Therefore, the flow does not play a role in the 
linear stability of a disordered suspension, implying that the force must be nonlinear in the local alignment 
$\bP$ (SI Text). We take the lowest nonlinear order permitted by symmetry and write $\bF =  \Phi |\bP|^2 \bP$. The proportionality constant $\Phi$ encodes the force per unit area generated by vesicles processing along perfectly aligned filaments, and thus includes all information relating to vesicle number, procession speed and binding rates.

\begin{figure*}
\centering
\includegraphics[width=\textwidth]{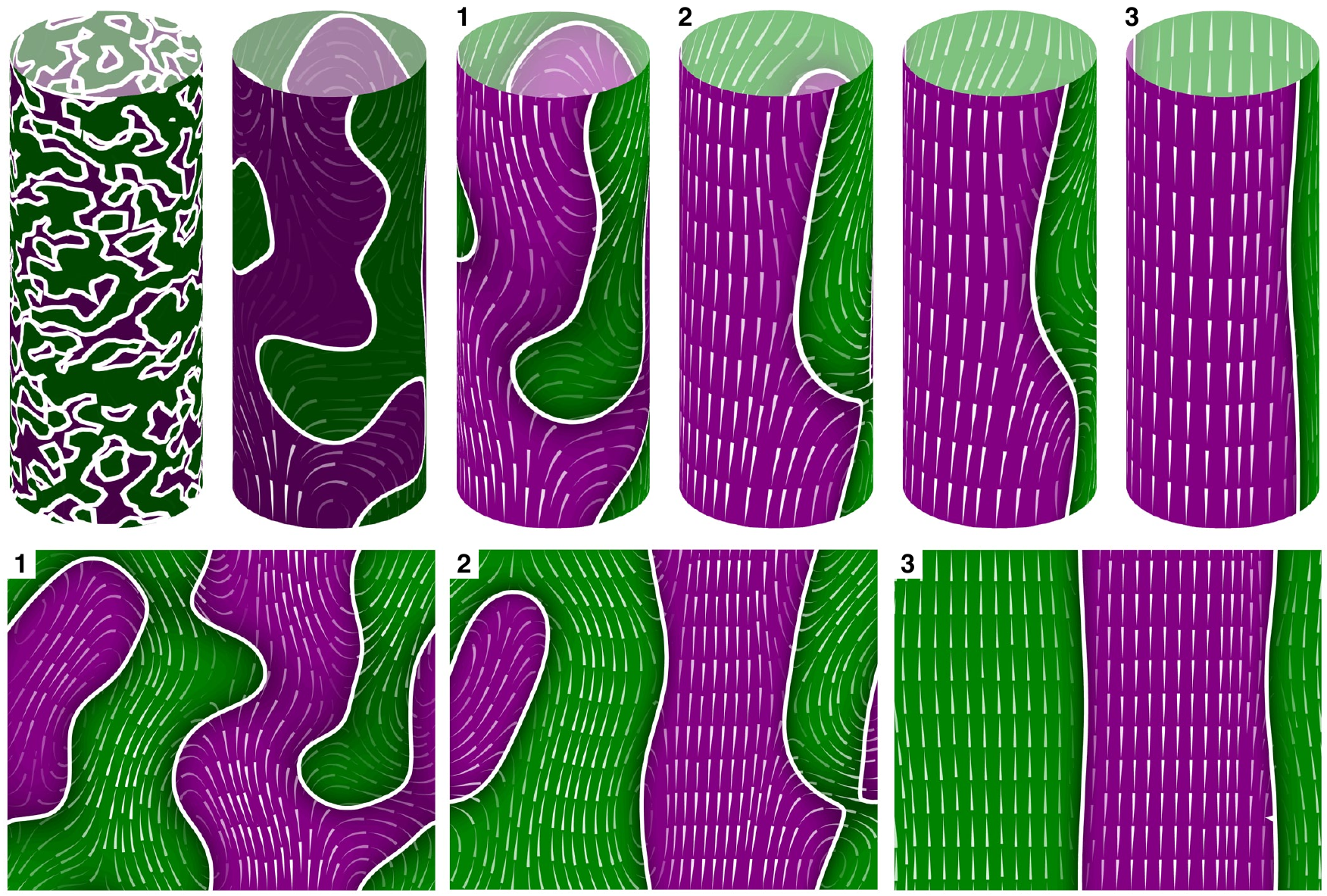}
\caption{Cyclosis developing in our model for $\alpha_p = 1$ and $\alpha_u = 0.5$, with other parameters fixed as in the text. Color-coding corresponds to the z-component of the order vector $\bP$, with purple for $P_z > 0$ and green for $P_z < 0$, darker for lower magnitude. White lines represent `indifferent zones' separating up- and down-streaming regions. Superimposed are streamlines of the cytoplasmic flow $\bu$ induced by the filament field $\bP$, where the flow is directed from the thin end to the thick end of the individual lines. Top: Time sequence of six frames, showing progression from random disorder through local order to complete steady cyclosis. Bottom: `Unwrapped' streaming patterns of the three indicated frames.
\label{fig:4}}
\end{figure*}

The last step is to non-dimensionalize all variables and coupling constants; we scale space by the radius $R$
and choose rescalings to eliminate the friction coefficient $\nu$ and the forcing constant $\Phi$ (SI Text).
The final dynamics for $\bP$ read
\begin{align*}
\frac{\partial \bP}{\partial t} + \epsilon \bu \cdot \nabla \bP &= d^{(s)} \nabla^2 \bP - d^{(r)} \bP + (\tI - \bP\bP)\cdot [\epsilon (\nabla\bu) \cdot \bP \\
& \qquad + \alpha_p \bP + \alpha_u \bu - \kappa (\bP\cdot\dhat)\dhat]
\end{align*}
where all coupling constants are now non-dimensional. The parameters $d^{(s)}$ and $d^{(r)}$ are spatial 
and rotational diffusion constants, respectively. The final hydrodynamics read
\begin{align*}
 -\mu \nabla^2 \bu + \bu + \Pi_0 \zhat + \nabla \Pi' = |\bP|^2 \bP,
\end{align*}
subject to
\begin{align*}
\nabla \cdot \bu = 0 \text{ and } \int_0^{2\pi} \int_0^\ell \bu\cdot\zhat \, dz \, d\theta = 0.
\end{align*}
All fields have periodic boundary conditions on $\theta = 0,2\pi$ and $z = 0,\ell$, where $\ell = L/R$ is the non-dimensional cell length.

\section{Results}
To test the model and understand more about the regimes of behavior it possesses, we ran numerical integrations of 
the system starting from initial conditions of randomly perturbed total filament disorder (see Materials and Methods). The 
parameters $\epsilon,d^{(s)},d^{(r)},\kappa,\mu$ were fixed at representative values 
$\epsilon=0.1$, $d^{(s)}=d^{(r)}=0.025$, $\kappa=0.5$ and $\mu=0.05$ in order to focus on the most important 
coupling constants $\alpha_p$ and $\alpha_u$. We also chose $\ell=5$ as an appropriate radius-to-length ratio 
for a young cell with observably organized actin cables~\cite{Wasteneys1996}.

\subsection{Time Progression}
Figure~\ref{fig:4} displays a typical time sequence for numerical integration of the model with an illustrative 
choice of the parameters $\alpha_p,\alpha_u$. The experimentally observed regeneration progression \cite{Foissner2000} is 
clearly reproduced as it moves from disorder, through small patches of locally ordered `streamlets' caused by 
spontaneous polarization, and settles into fully developed cyclosis as the passive and active reorienting effects of 
flow reach full potency. While the polar flow alignment is important for establishing the global streaming pattern,
it is worth remarking that the traditional nematic shear alignment, though damped by the restriction
factor $\epsilon$, still plays a role: it acts to smooth out
curved IZ boundaries between adjacent up- and down-streaming regions, where flow shear is high, into straight lines.
This is seen in the later stages of the time evolution in Fig.~\ref{fig:4}.

\subsection{Cyclosis Parameter Scan}

\begin{figure}
\centering
\includegraphics[width=\columnwidth]{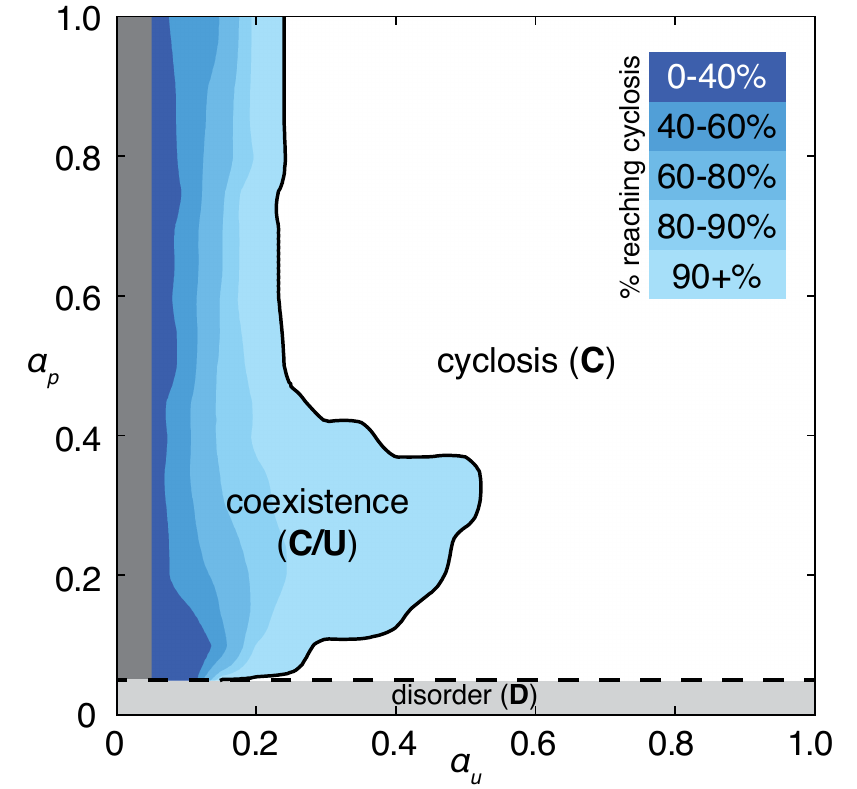}
\caption{Results of a numerical parameter sweep over $\alpha_p$ and $\alpha_u$ between $0.025$ and $1$, with other parameters held fixed as in the text. The entire set of $20 \times 20$ parameters was simulated for 50 different random initial conditions, uncovering regions of cyclosis/unidirectional bistability (C/U) and fully robust cyclosis (C). Color-coding in Region C/U denotes the percentage of the initial conditions that developed into steady cyclosis states at each pair of parameters. Region D, where $\alpha_p < d^{(r)}$, is linearly stable and develops no streaming.
\label{fig:5}}
\end{figure}

To gauge the model's robustness, we executed a numerical parameter sweep of non-zero $\alpha_p$ and $\alpha_u$ 
both between $0.025$ and $1$. The parameter space was divided up into a $20 \times 20$ grid, and the full set of 
parameters was simulated for each of 50 different instances of random initial conditions. Combined with a linear 
stability analysis (SI Text), this detects three distinct regimes (Fig.~\ref{fig:5}): disorder (D), 
cyclosis/unidirectional coexistence (C/U) and full cyclosis (C). Parameters in Region D have $\alpha_p$ too low for 
spontaneous order to flourish, where the disordered state $\bP=0$ is linearly stable.
In Region C/U, the system may evolve into either the cyclosis state or the unidirectional state $\bP = P_0 \zhat$ (where $P_0$ is a constant). Which steady state is reached is dependent on the precise nature of the initial conditions, a consequence of
co-existence of basins of attraction in the neighborhood of the Fourier space origin.
Parameters in Region C always developed the desired cyclosis 
solutions for all sets of initial conditions, indicating that the basin of attraction of the unidirectional state is now negligible.
These regimes can be understood further by considering the simplified one-dimensional regime of $z$-independent and $z$-parallel alignment, i.e. $\bP = P(\theta) \zhat$; see the SI Text.

\subsection{Flow-Stabilized Pattern Formation}
The one-dimensional simplification also allows us to consider the model in the context of pattern formation theory. When $\alpha_u=0$ the system assumes non-conservative Allen-Cahn form,
for which it can be shown that a spatially oscillating state is always unstable; consequently, fronts separating regions of upward and downward alignment
will always eventually coalesce~\cite{Pismen}.
However, $\alpha_u > 0$ introduces a difficult non-local term which cannot be written as the gradient of some free energy functional.
In fact, for $\alpha_u$ above a small threshold, cyclosis solutions are stabilized by global flow alignment effects.
The exact threshold as a function of $\alpha_p$ can be determined by fixing $\alpha_p$ and varying $\alpha_u$ in order to compute trajectories of eigenvalues of the
non-local Sturm-Liouville problem induced by linear stability analysis. This is discussed further in the SI Text.

\section{Discussion}

We have demonstrated the capability of this model for simulating the emergence of Characean streaming. For the computational study, we focussed on the roles of the polarization and flow-coupling constants $\alpha_p$ and $\alpha_u$. Perhaps the next most important parameter is the directional repulsion $\kappa$, which we have not investigated in detail here. The value we chose, $\kappa=0.5$, is a nice intermediate providing a definite preference for upwards orientation, but not so strong as to completely dominate the early-time dynamics. On the other hand, if $\kappa$ is too small, then little or no directional effects are felt, and the dynamics may be either unsteady or settle into circumferential flows.

In fact, studies of poor, malformed streaming induced \textit{in vivo} hint at the true effects of removing directional preference. Conversely, a suitable model for streaming should be able to reproduce observed pathologies and experimental disruptions, at least qualitatively, and it is these we now briefly discuss. 
Total cable fragmentation~\cite{Foissner2000, Foissner2007, Ryu1995} has already been mentioned, and directly corresponds to setting $\alpha_p = 0$ to inhibit bundling effects. There is also the possibility of atypical reorganization of the actin cables; to our knowledge, such behavior has not yet been induced in Characean cells. Nevertheless, it 
has been documented in other types of orderly streaming plant cells, with which a direct comparison can be made 
owing to the generality of this approach. On inhibiting the binding of the cell membrane to the cell wall, actin cables 
in {\it Vallisneria} mesophyll cells can reorient from longitudinal to circumferential, running around the long  
axis~\cite{Masuda1991}. Similar reorientation can be induced in {\it Lilium} 
pollen tubes, where actin cables reorient circumferentially in regions lacking longitudinal microtubules~\cite{Foissner2002}. Both studies 
supply evidence for the restriction and repulsive direction components of our model. Even though these atypical 
streaming arrangements were not induced during cell development, such a configuration can be reproduced by 
choosing parameters appropriately (SI Text and Fig.~S4).

We have considered the bases of $\alpha_p$ (bundling), $\kappa$ (geometric reorientation) and $\epsilon$ (restriction), 
but how might one experimentally account for the polar alignment coefficient $\alpha_u$? Setting 
$\alpha_u \approx 0$ corresponds to inhibiting the ability of myosin to walk on actin. In the Characeae one would 
also have to disorganize the filament network~\cite{Foissner2000} to observe its effect from a base state. 
We would then expect a unidirectional streaming pattern to emerge, or at least a metastable cyclosis with unequal 
widths of opposing streams, as the filaments bundle. Such an experiment would be an exciting test of the principal ideas on which this conceptually simple yet powerful self-organization model has been built.

\pagebreak



\begin{materials}
\fontsize{7pt}{9pt}\selectfont
\materialfont
\section{Simulation Method}
Simulations were performed using spectral methods, evaluating gradients and evolving in time in Fourier space and moving into physical space to evaluate nonlinearities. The periodic boundary conditions were then implicit in choosing a Fourier basis. Explicit 4th-order Runge-Kutta integration was used for time evolution.

\section{Initial Conditions}
For all simulations, the initial conditions were of total filament disorder as follows.
Suppose in a small area $dA$ there are $N \gg 1$ filaments whose orientations $\bP_i$, $1 \leq i \leq N$, are uniformly distributed, i.e. $\bP_i = (\cos \Theta_i, \sin \Theta_i)$ where $\Theta_i \sim \mathcal{U}[0,2\pi)$. Then since the $\{\Theta_i\}$ are independent and identically distributed random variables, the Central Limit Theorem applies to $\bP = \frac{1}{N} \sum_i \bP_i$, giving that $P_{\theta},P_z \sim \mathcal{N}(\mu, \sigma^2/N)$
with $\mu = \mathbb{E}(\cos \Theta_i) = \mathbb{E}(\sin \Theta_i) = 0$ and $\sigma^2 = \text{Var}(\cos \Theta_i) = \text{Var}(\sin \Theta_i) = 1/2$. Therefore, to model an initial distribution of $N \gg 1$ filaments per grid square, initial conditions are drawn as $P_{\theta},P_z \sim \mathcal{N}\left(0, \tfrac{1}{2N}\right)$.
$N = 1000$ was taken for all simulations.
\end{materials}


\begin{acknowledgments}
We thank J. Dunkel, A. Honerkamp-Smith, P. Khuc Trong, M. Polin and H. Wioland for discussions. This work was supported by the Engineering and Physical Sciences Research Council (EPSRC) and European Research Council (ERC) Advanced Investigator Grant 247333.
\end{acknowledgments}






\begin{thebibliography}{10}

\bibitem{KhucTrong2012}
Khuc~Trong P, Guck J, Goldstein RE
\newblock (2012) Coupling of active motion and advection shapes intracellular
  cargo transport.
\newblock \emph{Phys. Rev. Lett.} 109:28104.

\bibitem{Ganguly2012}
Ganguly S, Williams LS, Palacios IM, Goldstein RE
\newblock (2012) Cytoplasmic streaming in {D}rosophila oocytes varies with
  kinesin activity and correlates with the microtubule cytoskeleton
  architecture.
\newblock \emph{Proc. Natl. Acad. Sci. U.S.A.} 109:15109--15114.

\bibitem{Goehring2011}
Goehring NW, {et~al.}
\newblock (2011) Polarization of {PAR} proteins by advective triggering of a
  pattern-forming system.
\newblock \emph{Science} 334:1137.

\bibitem{Allen1978}
Allen NS, Allen RD
\newblock (1978) Cytoplasmic streaming in green plants.
\newblock \emph{Annu. Rev. Biophys. Bioeng.} 7:497--526.

\bibitem{Taylor1997}
Taylor LP, Hepler PK
\newblock (1997) Pollen germination and tube growth.
\newblock \emph{Annu. Rev. Plant Biol.} 48:461--491.

\bibitem{Hepler2001}
Hepler PK, Vidali L, Cheung AY
\newblock (2001) Polarized cell growth in higher plants.
\newblock \emph{Annu. Rev. Cell Dev. Biol.} 17:159--187.

\bibitem{Howard2011}
Howard J, Grill SW, Bois JS
\newblock (2011) Turing's next steps: the mechanochemical basis of
  morphogenesis.
\newblock \emph{Nature Rev. Mol. Cell Biol.} 12:392--398.

\bibitem{Kamiya1959}
Kamiya N
\newblock (1959) \emph{Protoplasmic Streaming}
\newblock (Springer-Verlag, Vienna).

\bibitem{Kruse2004}
Kruse K, Joanny JF, J{\"u}licher F, Prost J, Sekimoto K
\newblock (2004) Asters, vortices, and rotating spirals in active gels of polar
  filaments.
\newblock \emph{Phys. Rev. Lett.} 92:78101.

\bibitem{Allard2010}
Allard JF, Wasteneys GO, Cytrynbaum EN
\newblock (2010) Mechanisms of self-organization of cortical microtubules in plants
  revealed by computational simulations.
\newblock \emph{Mol. Biol. Cell} 21:278--286.

\bibitem{Surrey2001}
Surrey T, N{\'e}d{\'e}lec F, Leibler S, Karsenti E
\newblock (2001) Physical properties determining self-organization of motors
  and microtubules.
\newblock \emph{Science} 292:1167--1171.

\bibitem{Narayan2007}
Narayan V, Ramaswamy S, Menon N
\newblock (2007) Long-lived giant number fluctuations in a swarming granular
  nematic.
\newblock \emph{Science} 317:105--108.

\bibitem{Schaller2010}
Schaller V, Weber C, Semmrich C, Frey E, Bausch AR
\newblock (2010) Polar patterns of driven filaments.
\newblock \emph{Nature} 467:73--77.

\bibitem{Sanchez2012}
Sanchez T, Chen DTN, DeCamp SJ, Heymann M, Dogic Z
\newblock (2012) Spontaneous motion in hierarchically assembled active matter.
\newblock \emph{Nature} 491:431--434.

\bibitem{Zumdieck2005}
Zumdieck A, {et~al.}
\newblock (2005) Continuum description of the cytoskeleton: ring formation in the cell cortex.
\newblock \emph{Phys. Rev. Lett.} 95:258103.

\bibitem{Srivastava2013}
Srivastava P, Shlomovitz R, Gov NS, Rao M
\newblock (2013) Patterning of polar active filaments on a tense cylindrical membrane.
\newblock \emph{Phys. Rev. Lett.} 110:168104.

\bibitem{Corti1774}
Corti B
\newblock (1774) \emph{Osservazione Microscopiche sulla Tremella e sulla
  Circolazione del Fluido in una Pianta Acquajuola}
\newblock (Appresso Giuseppe Rocchi, Lucca, Italy).

\bibitem{Nothnagel1982}
Nothnagel EA, Webb WW
\newblock (1982) Hydrodynamic models of viscous coupling between motile myosin
  and endoplasm in characean algae.
\newblock \emph{J. Cell Biol.} 94:444--454.

\bibitem{Kachar1988}
Kachar B, Reese TS
\newblock (1988) The mechanism of cytoplasmic streaming in characean algal
  cells: sliding of endoplasmic reticulum along actin filaments.
\newblock \emph{J. Cell Biol.} 106:1545--1552.

\bibitem{Williamson1993}
Williamson RE
\newblock (1993) Organelle movements.
\newblock \emph{Annu. Rev. Plant Biol.} 44:181--202.

\bibitem{Sheetz1983}
Sheetz MP, Spudich JA
\newblock (1983) Movement of myosin-coated fluorescent beads on actin cables in
  vitro.
\newblock \emph{Nature} 303:31.

\bibitem{Shimmen2007}
Shimmen T
\newblock (2007) The sliding theory of cytoplasmic streaming: fifty years of
  progress.
\newblock \emph{J. Plant Res.} 120:31--43.

\bibitem{Kersey1976a}
Kersey YM, Hepler PK, Palevitz BA, Wessells NK
\newblock (1976) Polarity of actin filaments in characean algae.
\newblock \emph{Proc. Natl. Acad. Sci. U.S.A.} 73:165--167.

\bibitem{Kersey1976b}
Kersey YM, Wessells NK
\newblock (1976) Localization of actin filaments in internodal cells of
  characean algae. {A} scanning and transmission electron microscope study.
\newblock \emph{J. Cell Biol.} 68:264--275.

\bibitem{Goldstein2008}
Goldstein RE, Tuval I, van~de Meent JW
\newblock (2008) Microfluidics of cytoplasmic streaming and its implications
  for intracellular transport.
\newblock \emph{Proc. Natl. Acad. Sci. U.S.A.} 105:3663--3667.

\bibitem{vandeMeent2008}
van~de Meent JW, Tuval I, Goldstein RE
\newblock (2008) Nature's microfluidic transporter: rotational cytoplasmic
  streaming at high {P}{\'e}clet numbers.
\newblock \emph{Phys. Rev. Lett.} 101:178102.

\bibitem{vandeMeent2010}
van~de Meent JW, Sederman AJ, Gladden LF, Goldstein RE
\newblock (2010) Measurement of cytoplasmic streaming in single plant cells by
  magnetic resonance velocimetry.
\newblock \emph{J. Fluid Mech.} 642:5--14.

\bibitem{Foissner2000}
Foissner I, Wasteneys GO
\newblock (2000) Microtubule disassembly enhances reversible
  cytochalasin-dependent disruption of actin bundles in characean internodes.
\newblock \emph{Protoplasma} 214:33--44.

\bibitem{Jarosch1956}
Jarosch R
\newblock (1956) Die {I}mpulsrichtungs{\"a}nderungen bei der {I}nduktion der
  {P}rotoplasmastr{\"o}mung [The changes in the direction of momentum during the induction of
protoplasmic streaming].
\newblock \emph{Protoplasma} 47:478--486. German

\bibitem{Ryu1995}
Ryu JH, Takagi S, Nagai R
\newblock (1995) Stationary organization of the actin cytoskeleton in
  {V}allisneria: the role of stable microfilaments at the end walls.
\newblock \emph{J. Cell Sci.} 108:1531--1539.

\bibitem{Foissner1996}
Foissner I, Lichtscheidl IK, Wasteneys GO
\newblock (1996) Actin-based vesicle dynamics and exocytosis during wound wall
  formation in characean internodal cells.
\newblock \emph{Cell Motil. Cytoskel.} 35:35--48.

\bibitem{Foissner1997}
Foissner I, Wasteneys GO
\newblock (1997) A cytochalasin-sensitive actin filament meshwork is a
  prerequisite for local wound wall deposition in {N}itella internodal cells.
\newblock \emph{Protoplasma} 200:17--30.

\bibitem{Yotsuyanagi1953a}
Yotsuyanagi Y
\newblock (1953) Recherches sur les ph{\'e}nomen{\`e}s moteurs dans les
  fragments de protoplasme isol{\'e}s. {I}. {M}ouvement rotatoire et le
  processus de son apparition [Research on the active phenomenon in isolated cytoplasm fragments. I.
Rotatory movement and the process of emergence].
\newblock \emph{Cytologia} 18:146--156. French

\bibitem{Yotsuyanagi1953b}
Yotsuyanagi Y
\newblock (1953) Recherches sur les ph{\'e}nom{\`e}nes moteurs dans les
  fragments de protoplasme isol{\'e}s. {II}. {M}ouvements divers
  d{\'e}termin{\'e}s par la condition de milieu [Research on the active phenomenon in isolated cytoplasm fragments. II. Diverse movements depending on the medium conditions].
\newblock \emph{Cytologia} 18:202--217. French

\bibitem{Woodhouse2012b}
Woodhouse FG, Goldstein RE
\newblock (2012) Spontaneous circulation of confined active suspensions.
\newblock \emph{Phys. Rev. Lett.} 109:168105.

\bibitem{Wasteneys1996}
Wasteneys GO, Collings DA, Gunning BES, Hepler PK, Menzel D
\newblock (1996) Actin in living and fixed characean internodal cells:
  identification of a cortical array of fine actin strands and chloroplast
  actin rings.
\newblock \emph{Protoplasma} 190:25--38.

\bibitem{Green1962}
Green PB
\newblock (1962) Mechanism for plant cellular morphogenesis.
\newblock \emph{Science} 138:1404.

\bibitem{Green1954}
Green PB
\newblock (1954) The spiral growth pattern of the cell wall in {N}itella
  axillaris.
\newblock \emph{Am. J. Bot.} 41:403--409.

\bibitem{Woodhouse2012a}
Woodhouse FG, Goldstein RE
\newblock (2012) Shear-driven circulation patterns in lipid membrane vesicles.
\newblock \emph{J. Fluid Mech.} 705:165--175.

\bibitem{Masuda1991}
Masuda Y, Takagi S, Nagai R
\newblock (1991) Protease-sensitive anchoring of microfilament bundles provides
  tracks for cytoplasmic streaming in {V}allisneria.
\newblock \emph{Protoplasma} 162:151--159.

\bibitem{Foissner2002}
Foissner I, Grolig F, Obermeyer G
\newblock (2002) Reversible protein phosphorylation regulates the dynamic
  organization of the pollen tube cytoskeleton: effects of calyculin {A} and
  okadaic acid.
\newblock \emph{Protoplasma} 220:1--15.

\bibitem{Tominaga1997}
Tominaga M, Morita K, Sonobe S, Yokota E, Shimmen T
\newblock (1997) Microtubules regulate the organization of actin filaments at
  the cortical region in root hair cells of {H}ydrocharis.
\newblock \emph{Protoplasma} 199:83--92.

\bibitem{Brotto2013}
Brotto T, Caussin JB, Lauga E, Bartolo D
\newblock (2013) Hydrodynamics of confined active fluids.
\newblock \emph{Phys. Rev. Lett.} 110:038101.

\bibitem{Parmeggiani2001}
Parmeggiani A, J{\"u}licher F, Peliti L, Prost J
\newblock (2001) Detachment of molecular motors under tangential loading.
\newblock \emph{Europhys. Lett.} 56:603.

\bibitem{Shimmen2004}
Shimmen T, Yokota E
\newblock (2004) Cytoplasmic streaming in plants.
\newblock \emph{Curr. Opin. Cell Biol.} 16:68--72.

\bibitem{Kimura2003}
Kimura Y, Toyoshima N, Hirakawa N, Okamoto K, Ishijima A
\newblock (2003) A kinetic mechanism for the fast movement of {C}hara myosin.
\newblock \emph{J. Mol. Biol.} 328:939--950.

\bibitem{Kantsler2012}
Kantsler V, Goldstein RE
\newblock (2012) Fluctuations, dynamics, and the stretch-coil transition of
  single actin filaments in extensional flows.
\newblock \emph{Phys. Rev. Lett.} 108:38103.

\bibitem{Saintillan2008}
Saintillan D, Shelley MJ
\newblock (2008) Instabilities, pattern formation, and mixing in active
  suspensions.
\newblock \emph{Phys. Fluids} 20:123304.

\bibitem{Pismen}
Pismen LM
\newblock (2006) \emph{Patterns and Interfaces in Dissipative Dynamics}
\newblock (Springer, Berlin).

\bibitem{Foissner2007}
Foissner I, Wasteneys GO
\newblock (2007) Wide-ranging effects of eight cytochalasins and latrunculin
  {A} and {B} on intracellular motility and actin filament reorganization in
  characean internodal cells.
\newblock \emph{Plant Cell Physiol.} 48:585--597.

\end{thebibliography}

\end{article}

\end{document}